\documentclass[sigconf]{acmart}

\AtBeginDocument{%
  \providecommand\BibTeX{{%
    \normalfont B\kern-0.5em{\scshape i\kern-0.25em b}\kern-0.8em\TeX}}}

\copyrightyear{2024}
\acmYear{2024}
\setcopyright{rightsretained}
\acmConference[CHI EA '24]{Extended Abstracts of the CHI Conference on Human Factors in Computing Systems}{May 11--16, 2024}{Honolulu, HI, USA}
\acmBooktitle{Extended Abstracts of the CHI Conference on Human Factors in Computing Systems (CHI EA '24), May 11--16, 2024, Honolulu, HI, USA}
\acmDOI{10.1145/3613905.3651122}
\acmISBN{979-8-4007-0331-7/24/05}

\usepackage{subfigure}
\usepackage{framed}

\begin{document}

\title{Empowering Personalized Learning through a Conversation-based Tutoring System with Student Modeling}

\author{Minju Park}
\email{minju.park@riiid.co}
\orcid{0000-0002-2588-2674}
\affiliation{
  \institution{Riiid AI Research}
  \city{Seoul}
  \country{Republic of Korea}
}
\author{Sojung Kim}
\email{sojung.kim@riiid.co}
\orcid{0000-0003-4585-3587}
\affiliation{
  \institution{Riiid AI Research}
  \city{Seoul}
  \country{Republic of Korea}
}
\author{Seunghyun Lee}
\email{seunghyun.lee@riiid.co}
\orcid{0009-0004-2082-9091}
\affiliation{
  \institution{Riiid AI Research}
  \city{Seoul}
  \country{Republic of Korea}
}
\author{Soonwoo Kwon}
\email{soonwoo.kwon@riiid.co}
\orcid{0000-0003-1764-3128}
\affiliation{
  \institution{Riiid AI Research}
  \city{Seoul}
  \country{Republic of Korea}
}
\author{Kyuseok Kim}
\email{kimkyu80@gmail.com}
\orcid{0009-0004-2369-6335}
\affiliation{
  \institution{Riiid AI Research}
  \city{Seoul}
  \country{Republic of Korea}
}

\begin{abstract}
    As the recent Large Language Models(LLM's) become increasingly competent in zero-shot and few-shot reasoning across various domains, educators are showing a growing interest in leveraging these LLM's in conversation-based tutoring systems. However, building a conversation-based personalized tutoring system poses considerable challenges in accurately assessing the student and strategically incorporating the assessment into teaching within the conversation. In this paper, we discuss design considerations for a personalized tutoring system that involves the following two key components: (1) a student modeling with diagnostic components, and (2) a conversation-based tutor utilizing LLM with prompt engineering that incorporates student assessment outcomes and various instructional strategies. Based on these design considerations, we created a proof-of-concept tutoring system focused on personalization and tested it with 20 participants. The results substantiate that our system's framework facilitates personalization, with particular emphasis on the elements constituting student modeling. A web demo of our system is available at \url{http://rlearning-its.com}.
\end{abstract}


\maketitle

\section{Introduction}

Large Language Models (LLM’s) such as GPT~\cite{brown2020language,openai2023gpt4} and PaLM~\cite{chowdhery2022palm}are excelling in diverse tasks, including common sense and knowledge reasoning, to mathematical, symbolic and visuo-linguistic tasks.  
Surprisingly, these models are shown to demonstrate zero-shot and few-shot reasoning capacities somewhat akin to general artificial intelligence~\cite{bubeck2023sparks}, despite being trained with a simple autoregressive text generation objective. 

Accordingly, a growing number of educators are showing interest in leveraging these LLM's to design an all-time available, personalized tutoring system\footnote{See also \url{https://www.khanacademy.org/} and \url{https://www.duolingo.com/} for product examples of interactive, personalized education platforms utilizing LLM's.}~\cite{KASNECI2023102274, macina-etal-2023-opportunities}.
For example, prior works have developed LLM-driven conversation-based tutors in various skill domains, including argumentation~\cite{ArgueTutor2021}, reading comprehension~\cite{DIRECT2023}, mathematical reasoning~\cite{macina2023mathdial}, and language learning~\cite{UALLchatbot2023}. 
However, most prior works lack a systematic student assessment, relying completely on text inputs implicitly.
Such absence of diagnostic capabilities may limit the level of achievable personalization. 

To address the aforementioned issue, we discuss the design considerations for a conversation-based personalized tutoring system.
Specifically, we first focus on the diagnostic components of the tutoring system, where these components allow the LLM to incorporate multi-faceted student assessment results via prompting. 
We consider three assessment criteria: cognitive state, affective state, and learning style.

Next, we focus on implementing personalized instructional strategies, where the LLM is prompted to provide user-tailored tutoring by incorporating diagnostic results. 
Specifically, we equip the system with various personalized instructional strategies such as adaptive exercise selection, interventional messages tailored to individual cognitive and affective states, and teaching customized to learning style. 
Overall, the resulting design is a versatile personalized tutoring system that utilizes a LLM in a plug-and-play fashion.
We believe such system can be implemented across various skill domains through the use of domain-specific prompts.

Building upon these design considerations, we introduce our proof-of-concept implementation of a personalized tutoring system.\footnote{For readers interested in experiencing the system firsthand, a web demo is accessible at \url{http://rlearning-its.com}.}
Our system teaches three English writing concepts of ``Pronouns'', ``Punctuation'', and ``Transitions'' from the SAT Writing test.  
\begin{figure}[t]
\centering
\includegraphics[width=0.5\textwidth]{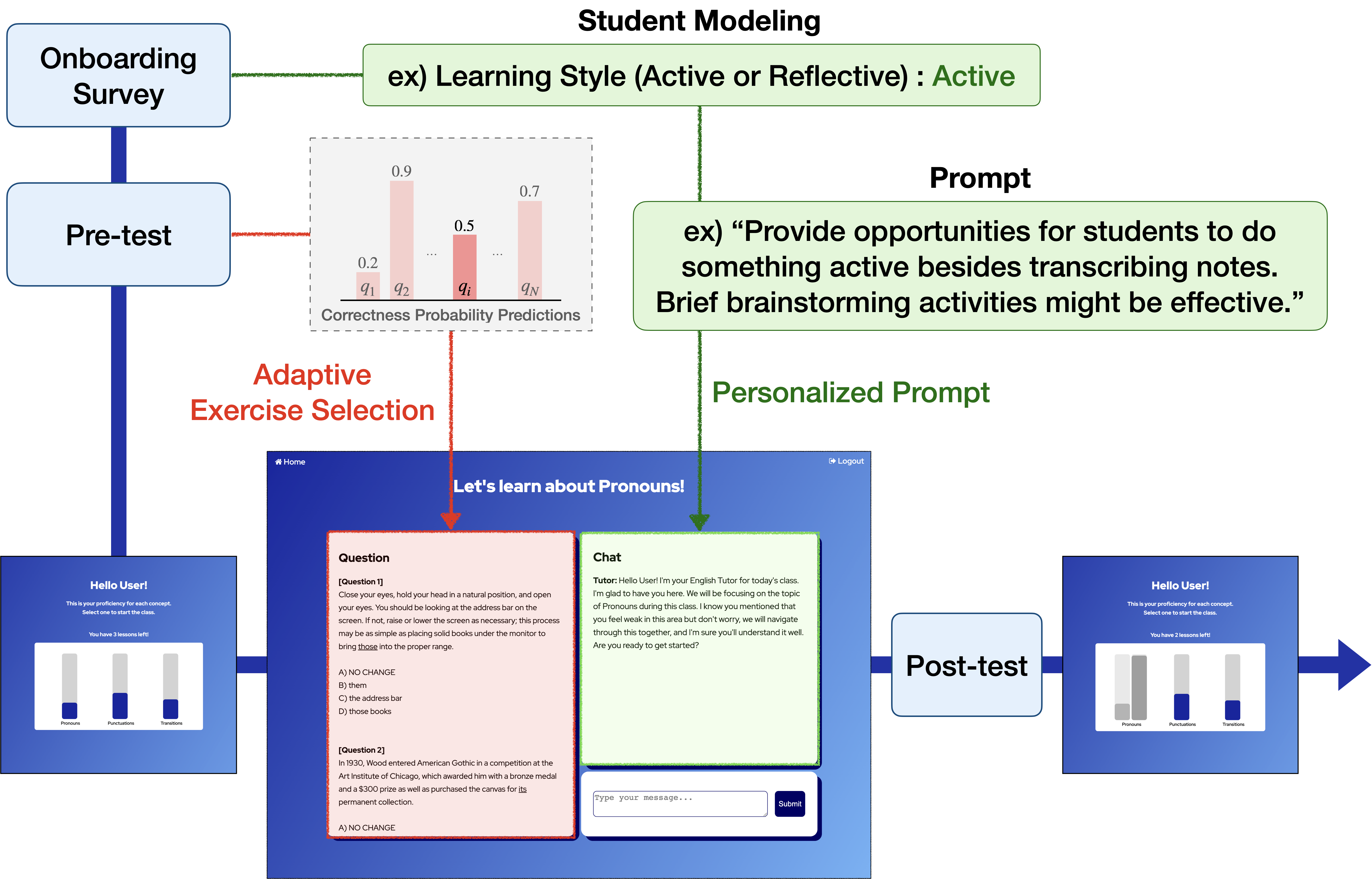}
\caption{Overview of our proof-of-concept personalized tutoring system}
\label{fig:overview}
\end{figure}
As depicted in Figure~\ref{fig:overview}, the user progresses through an onboarding survey, pre-test, tutoring session, and post-test for each concept, with subsequent concepts following the same pattern. 
Throughout the process, the student states undergo assessment and are subsequently applied to the main tutoring session.

In order to ascertain the system's ability in implementing personalization within the tutoring process, we executed tutoring sessions with the participation of 20 individuals.
We provide comprehensive instances of how our system applies personalization in student modeling.
Additionally, we offer insights and discussions stemming from the results.

We highlight the main contributions of our paper below:
\begin{itemize}
    \item We discuss important design considerations for a personalized tutoring system that involves (1) diagnostic components that support student assessment; and (2) a conversation-based tutor utilizing LLM prompted to incorporate student assessment and instructional strategies into teaching.
    \item Based on the design considerations, we implement a simple proof-of-concept tutoring system.
    \item We closely examined interactions between the system and actual learners to assess its alignment with our design objectives. This process provided critical insights into potential improvements and identified specific challenges, thereby inviting the HCI community to address these issues and advance the field of personalized educational technology.
\end{itemize}

\section{Related Work}

\subsection{Intelligent Tutoring Systems}
Intelligent Tutoring Systems (ITS) have been a subject of extensive research for providing immediate, personalized one-to-one tutoring experiences to learners through computer-based educational technologies~\cite{advances-its, nye2014autotutor}.
As LLM's such as GPT and PaLM have revolutionized the landscape of natural language generation, it has also extended its potential to facilitate natural language-based ITS~\cite{Autotutor2005}, which serves as an AI-equipped tutoring agent that engages in dialogues with the students.
Such integration of natural language interface holds the potential to enhance the personalization, interactivity, and adaptability of learning experiences within the context of ITS. 

Recent advances in conversation-based tutoring systems utilizing LLM's have showcased their transformative potential~\cite{KASNECI2023102274}. 
For instance, ArgueTutor~\cite{ArgueTutor2021} offers adaptive tutoring for argumentation skills, DIRECT~\cite{DIRECT2023} specializes in reading comprehension tutoring systems, MATHDIAL~\cite{macina2023mathdial} focuses on math reasoning problems, and the work by~\cite{UALLchatbot2023} enhances language learning experiences. 
Moreover, prompt engineering-based tutoring systems have broadened the horizons of possibilities~\cite{chen2023gptutor, RECIPE2023}. 

However, the mentioned tutoring systems lack the explicit incorporation of student modeling, relying instead on text-based cues to track student progress. 
The potential downside to this approach is that these systems may struggle with insufficient diagnostic capabilities resulting from the gap in explicit student modeling. 
This absence of diagnostic capabilities can significantly impact the level of achievable personalization, particularly in terms of selecting adaptive tutoring strategies based on student assessment. 
We expect this limitation to be especially profound for a tutoring system that completely relies on in-context learning (zero- or few-shot) of LLMs, due to lack of task-specific fine-tuning. 

Addressing such limitations requires the integration of extensive student modeling from various perspectives, e.g., cognitive and affective states, learning style, etc.  
Moreover, the development of LLM prompt-based systems should consider incorporating a system structure where prompts are adaptively updated based on student modeling. 
This dual approach of refining student modeling and designing user-adaptive prompts for conversation-based tutoring system can collectively contribute to the adaptability and personalization of the learning process.

\subsection{Cognitive Diagnostic Modeling}
Cognitive Diagnosis Modeling (CDM) refers to the task of assessing a given examinee's knowledge proficiency based on test results. 
Here, a test is usually comprised of questions whose correctness can be modeled by a binary random variable.
Since accurate assessment can result in the efficient selection of learning materials, CDM's are often employed for item recommendations within e-learning environments, including ITS~\cite{wong2010learning}.

The most influential, traditional methods for CDM include Item Response Theory (IRT; \cite{embretson2013item}) and DINA~\cite{de2009dina}. 
Within this framework, we intend to employ CDM to assess students using the IRT model.
The outcome of the assessments would subsequently provide adaptive recommendations for exercises, thereby enhancing the personalization of our system.

\section{Design of Personalized Tutoring System}
In this section, we discuss design considerations for a personalized tutoring system that involves (1) multi-faceted student assessment, and (2) LLM prompt-based tutoring that incorporates student assessment outcomes and various instructional strategies.

To aid in understanding the context of these considerations, we first outline the user flow that has been designed to optimize their application within our entire approach.
The flow begins with an onboarding survey that introduces users to the system's features and establishes their initial student profile.
They then take a 15-question pre-test on SAT Writing concepts, followed by personalized tutoring based on the pre-test results.
After tutoring, students take a post-test to gauge their improvement, and this iterative process continues until all three knowledge concepts are covered.
For more in-depth information regarding the specifics of this user flow and its alignment with our design considerations, readers are encouraged to refer to the Appendix \ref{sec:appendix_userflow}.

\subsection{Student Assessment}
We first address student assessment criteria of our student modeling. 
They are strategically chosen not only for their pedagogical significance but also based on the capabilities of LLM, particularly GPT-4, to understand and operationalize these criteria. The system relies on three fundamental assessment criteria to shape its instructional strategies: (1) cognitive state, (2) affective state, and (3) learning style. Each criterion is discussed in detail below. For every criterion, we provide a definition, measurement methodology, and relevant learning science literature that underpins our approach.

\subsubsection{Cognitive State}

A cognitive state encompasses various statuses influencing an individual's abilities in learning, comprehension, problem-solving, and decision-making.
This includes, but is not limited to, factors such as attention, memory, problem-solving abilities, proficiency levels in particular knowledge domains, and metacognition. In the context of this study, we narrow our focus to those aspects of cognitive state that can be empirically assessed using our system, which incorporates both the CDM and GPT-4. Specifically, we examine:

\begin{itemize}
\item Proficiency levels in knowledge concepts, as quantified by IRT;
\item Metacognition, as captured through self-reported data on students' self-assessment of their grades and proficiency, which offers insights into their self-regulatory practices and self-awareness;
\item GPT-4's session-end summary, which provides an evaluation of a student's cognitive state based on their interactions during the tutoring session, and;
\item Learning Gain, as measured by the difference in proficiency between pre-test and post-test, facilitating insights into student progress as well as engagement levels, and shaping action items for subsequent tutoring sessions.
\end{itemize}

The importance of considering a student's cognitive state in personalized learning is well-documented in educational research. Proficiency levels in knowledge concepts are crucial for selecting learning materials that fall within the student's Zone of Proximal Development, optimizing the effectiveness of the educational experience~\cite{hung2001situated}. 
Furthermore, metacognitive knowledge and learning strategies have been found to influence learning outcomes independently of intelligence~\cite{Paul2002, Leutwyler2009, VEENMAN200489}.

\subsubsection{Affective State}

Affective state refers to the emotional and attitudinal aspects that influence a learner's engagement with their overall learning experience, encompassing elements such as motivation, interest, and self-efficacy. In this study, we aim to quantify the affective state through the following means, acknowledging that our diagnosis of the affective state is inherently limited.

\begin{itemize}
\item The discrepancy between proficiency levels determined by IRT models and those self-reported by the students. This metric helps identify gaps between perceived and actual skill levels, which can serve as indicators of emotional factors like self-efficacy, anxiety, or overconfidence.

\item GPT-4's session-end summary, which encapsulates key affective indicators including engagement level, motivation, and favorability toward the subject matter. These indicators are inferred from student dialogues and interactions that occur during the tutoring sessions.

\end{itemize}

Prior research has shown that emotional and attitudinal factors not only act as reliable predictors of academic success but also have a profound impact on the efficacy of teaching methodologies~\cite{Pekrun1992, Miriam2005,Rebecca2000,WIGFIELD200068}. 

\subsubsection{Learning Style}

Learning style denotes the preferred means by which an individual absorbs, processes, comprehends, and retains information~\cite{Felder1988,felder2002learning, el2019use}. 
We adopt the Felder and Silverman learning style model~\cite{felder2002learning}, which classifies learning styles into 16 categories, depending on how one prefers the modes of: Perception (sensory vs intuitive), Input (visual vs auditory), Processing (active vs reflective), and Understanding (sequential vs global).
As our system relies on a chat-based interface, the adaptation of teaching methods within the Input mode was constrained.
Therefore, our system takes into consideration three distinct modes: Perception, Processing, and Understanding.

In our system, learning styles are identified by:
\begin{itemize}
    \item Self-reported learning style, where students explicitly indicate their preferences across the three dimensions based on provided explanations. This data sets the initial conditions for personalized instruction, offering an individualized starting point for each student's educational journey.
    
    \item GPT-4's session-end summary, which tries to discern the learning styles demonstrated during the tutoring sessions by examining modes of interaction and question-answering patterns. Such insights are then used for continual adjustments in learning strategies, supplemented with comments.

\end{itemize}
Numerous studies suggest that understanding a student's learning style can profoundly enhance the adaptation of learning and teaching methods~\cite{Huang2012, el2019use, Assis2022, Lim2021}. 
We prioritize learning style as it provides a clear framework to refine our prompting algorithms, further individualizing the teaching experience.

The described student modeling will underpin the creation and design of the personalized tutoring that we’ll delve
into in the following section. 
We have designed GPT-4's session-end summary to incorporate specific action items related to instructional strategies based on the assessed status of the student. 
This integration is intentional, aiming to directly and explicitly feed strategic actions into GPT-4's system prompt for more targeted tutoring.

\subsection{LLM Prompt-Based Personalized Tutoring System }

\begin{figure*}[ht]
\centering
\includegraphics[width=0.8\textwidth]{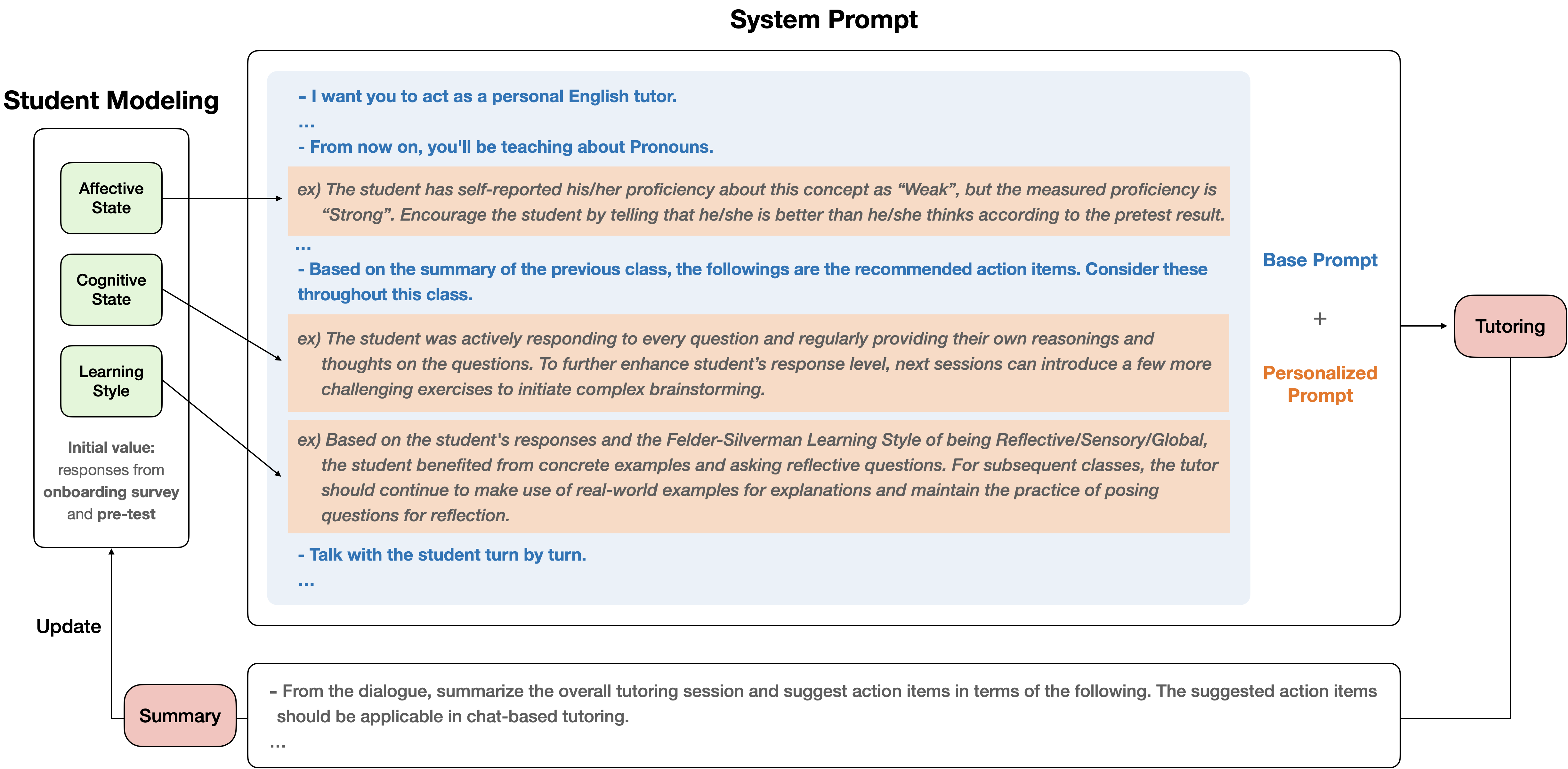}
\caption{Cyclical framework and structure of the system prompt in our proof-of-concept personalized tutoring system}
\label{fig:prompt}
\end{figure*}

In this section, we discuss the personalization strategies we have implemented in our proof-of-concept tutoring system.
With this user flow as our foundation, we then explore how to employ cognitive diagnostic modeling, namely IRT, and leverage the specific capabilities of GPT-4 in our proof-of-concept system to design prompts. 
The overall structure of prompt design within our system is illustrated in Figure~\ref{fig:prompt}.

\subsubsection{Adaptive Exercise Selection} \label{sec:item}

The first strategy we consider is the adaptive selection of learning materials for tutoring.
We used the one-dimensional 2PL IRT model for each knowledge component, where the probability of a correct response is modeled by:
\begin{align}
P(\textrm{correct}| a, d, \theta) = \frac1{1 + \exp(-a \cdot (\theta - d) )},  
\label{eq:irt}
\end{align}
with $a$, $d$ denoting the item's discrimination and difficulty parameters, respectively, and $\theta$ denoting the student's skill parameter.
Our primary focus is on selecting exercises that have a probability of a correct answer close to 0.5. This probability is determined using pre-test results and pre-computed item parameters.
We anticipate maximal learning gains when students are presented with items that are neither too easy nor too difficult~\cite{vygotsky1978mind}. Although we utilize this specific recommendation system in our approach, other systems could also be integrated into this framework.

Beginning by pre-determining the item parameters $a, d$ for every item related to the knowledge concept through IRT modeling (Eq.~\ref{eq:irt}), we compute the user's skill parameter $\theta$ based on the user's pre-test results. Then with the user's skill parameter available, we select the first three exercises whose difficulty parameter $d$ is closest to $\theta$.

The exercises we select, along with their correct answers and detailed explanations, are incorporated into the prompt. 
This allows the LLM tutor to effortlessly refer to them and collaboratively work through them with the student. 
At its core, our adaptive exercise selection is rooted in the student's cognitive state, as gauged by IRT, directing our choice of exercises for each tutoring topic.

\subsubsection{Prompt Design for Personalized Tutoring} \label{sec:prompt}
In our quest to enhance the efficacy of personalization through LLM prompt-based tutoring, we carefully design the prompt structure. 
Figure~\ref{fig:prompt} visualizes our approach: a cyclical framework intertwining student assessment and system prompt. 
The system prompt splits into two main facets: the base prompt and personalized prompt.
On the left side of Figure~\ref{fig:prompt}, the student modeling component informs the system prompt, supplying inputs for the personalized prompt. 
This refined system prompt subsequently guides the tutoring session. 
After the session, the process transitions to the summary, wherein a summary prompt is integrated to evaluate student interactions throughout the session. 
Feedback from this summary cycles back, updating the student assessment and reinforcing the continuous loop of our system. 

Our system employs a dual-prompt approach, comprising the base and personalized prompts. The base prompt provides the foundational guidelines that structure the tutoring interaction. It ensures that the tutoring session is interactive, student-centered, and adheres to effective pedagogical practices. An example of the base prompt is highlighted in blue in Figure~\ref{fig:prompt}.

On the other hand, the personalized prompt varies among students, adapting to each student's unique characteristics. It evolves based on individual student modeling, with its initial form shaped by the student states constructed immediately after the pre-test. The examples in Figure~\ref{fig:prompt}, highlighted in orange, demonstrate how prompts can be tailored to a student's cognitive state, affective state, and learning style.
The personalized prompt is designed to integrate information gathered from the pre-test and onboarding survey. For subsequent sessions, it also includes the information of the current student's state based on the summaries of prior sessions to ensure the instructions provided are relevant to the student's progress.
The summary prompt serves as a bridge between student interaction and GPT-4's assessment capabilities, distilling key insights from tutoring sessions. It is designed to extract session recaps, assess student proficiency, and suggest action items for subsequent tutoring. The assessment and suggested actions, derived from the summary, contribute to ongoing refinement of student modeling and the system prompt. This iterative process ensures the incorporation of action item recommendations for tutoring strategies in the system prompt from the second session onward. The detailed summary prompt is available in Appendix \ref{sec:appendix_summary_prompt}.

To offer a tangible insight into the application of our system, we have provided an example of the entire system prompt - used with one of the participants - in the Appendix \ref{sec:appendix_system_prompt}. 
Therein, the specifics of the system prompt and the method of integrating summaries are  elucidated.

\section{Results}
In this section, we present several observations from the examination of our proof-of-concept tutoring system with illustrative examples.
We conducted a recruitment of 20 individuals, who exhibited a diverse range of English proficiency levels, spanning from novices to advanced proficiency.
They engaged in the complete cycle of our tutoring system, and we conducted an analysis of the dialogues as a basis for our findings.
Table~\ref{tab:summary} provides an encompassing overview of our tutoring dataset.

\begin{table}[h]
\renewcommand*{\arraystretch}{1.3}
\begin{center}
\begin{tabular}{l|l}
\toprule
\multicolumn{1}{c|}{\textbf{Dataset Summary}} & \multicolumn{1}{c}{\textbf{Counts}} \\ \midrule
Dialogues & 74\\
Turns & 1470\\
Avg. Turns per dialogue & 19.87\\
Avg. Words per utterance (Tutor) & 72.45\\
Avg. Words per utterance (Student) & 3.9\\ \bottomrule
\end{tabular}
\end{center}
\caption{Tutoring dataset summary}
\label{tab:summary}
\end{table}

\begin{figure}[h]
    \centering
    \subfigure[]{
        \includegraphics[width=0.45\textwidth]{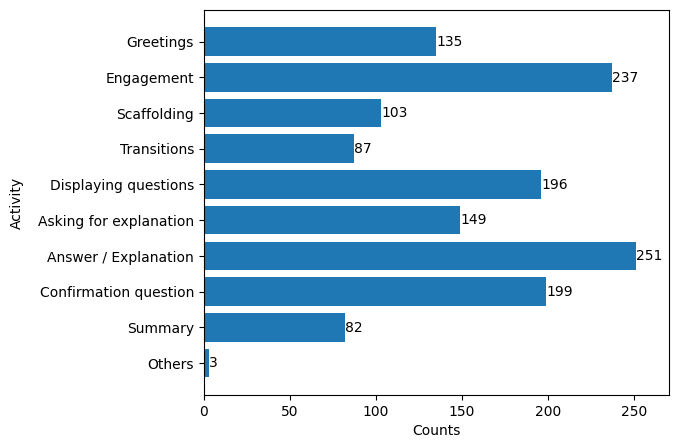}
        \label{fig:interventions_hist}
    }
    \subfigure[]{
        \includegraphics[width=0.45\textwidth]{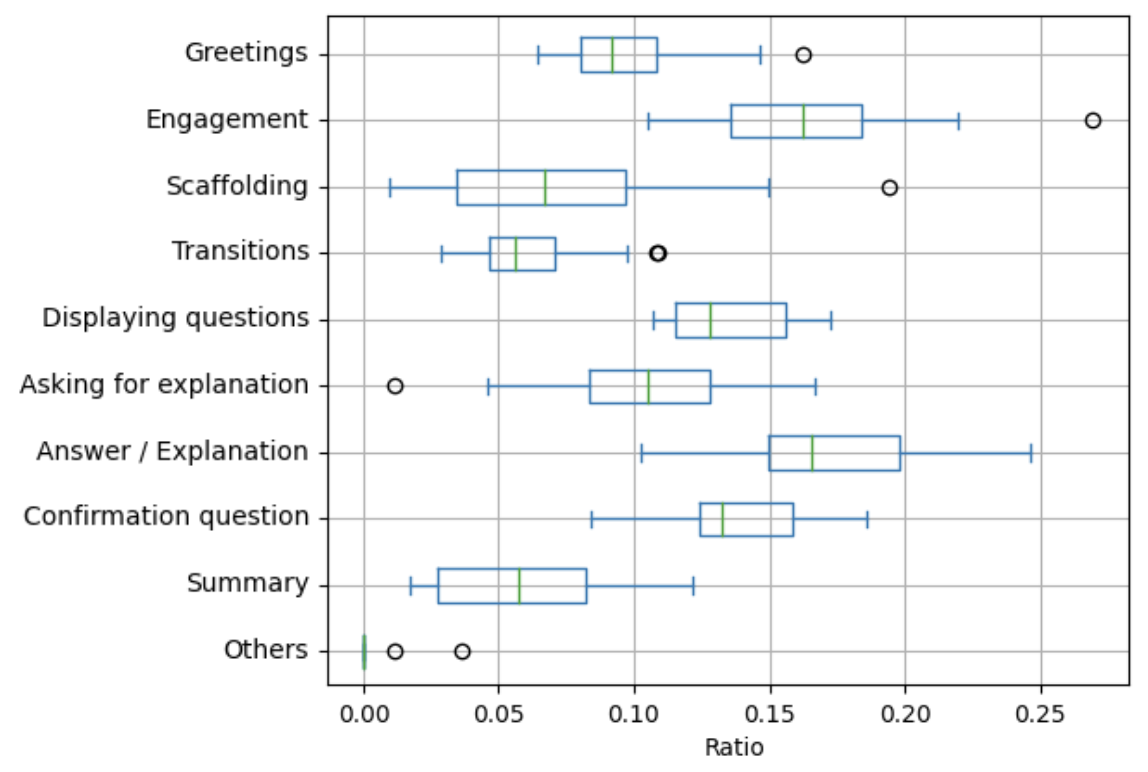}
        \label{fig:interventions_box}
    }
    \caption{(a) Counts of labeled tutor actions across entire dialogues. (b) Distribution of the ratio of each tutor action computed on a student-wise basis.}
    \label{fig:tutor_actions}
\end{figure}

We manually labeled each tutor's utterance to understand the dialogues' content and nature. 
We categorized the utterances into 10 actions, including Greetings, Engagement, and Summary, based on a prior study~\cite{stasaski-etal-2020-cima}, with slight modifications to fit our context. The distribution of labeled tutor actions is shown in Figure~\ref{fig:interventions_hist}.

\subsection{Observation}

From Figure~\ref{fig:interventions_box}, we examined the varying proportions of tutor actions within a session across different students. The variation in the appearance rate of tutor actions such as Engagement, Scaffolding, Asking for explanation, and Summary within the students indicates that the tutoring system is successfully choosing teaching approaches that are well-suited to each individual student.

The examples presented in tutoring dialogues further illustrate the occurrence of personalization within the context of tutoring sessions. These examples can be traced back to three foundational dimensions of student assessment: cognitive state, affective state, and learning style.

For our first example, we turn to how the system leverages student metacognition to offer personalized teaching techniques. In one particular scenario, a student's self-assessed proficiency was notably lower than what the student's pre-test results indicated. Recognizing this disparity, the tutoring system provides positive reinforcement by stating:
\begin{quote}
    ``\textit{It's interesting to note that based on your pre-test results, I see that your knowledge about punctuation is actually stronger than you think! That's a great start.}”
\end{quote}

Secondly, in a situation where the student has consistently shown high engagement, the GPT-4 has observed and suggested future action items as,
\begin{quote}
    ``\textit{Encouraging the student to go beyond the right answer and explain the reasoning behind their choices in their own words proved to be an effective strategy. For future sessions, continuous prompts to express thought processes or reasoning behind answers could further stimulate engagement and verify understanding.}''
\end{quote}
During the subsequent session, the tutor reflected its action items by saying, 
\begin{quote}
    ``\textit{That's correct! The correct answer is indeed A, ``NO CHANGE''. You realized that ``its'' was the right pronoun in this scenario. Could you elaborate a bit on what helped you understand that ``its'' should not be changed?}'' 
\end{quote}

Lastly, personalized tutoring strategies were implemented according to each student's individual learning style.
This approach was applied not only during the initial session, where tutoring strategies were explicitly tailored based on the reported learning style obtained from the onboarding survey, but also in subsequent sessions as the learning style was updated by the GPT-4.
For instance, GPT-4 analyzed the student's learning style and recommended strategies as 
\begin{quote}
    ``\textit{To cater to the student's global learning style, the tutor could use summarizations and reinforcements of lesson points throughout future sessions to highlight the overall context and relevance of the studied matter.}''
\end{quote}
after the first session.
During the following session, the tutor asked the student 
\begin{quote}
    ``\textit{To deepen your understanding, could you summarize what you've just learned from this question?}'',
\end{quote}
applying the recommended strategy.

Overall, the system successfully adjusted instructional strategies in a personalized manner.
Additionally, the GPT-4 demonstrated its effectiveness in analyzing and evaluating the student states based on the dialogues when necessary.

\subsection{Discussion}
While our system demonstrates the potential of LLMs in conversation-based tutoring, we also uncovered limitations. 
A primary concern is that precise student assessment doesn't always result in discernible, actionable tutoring strategies, possibly due to our system's focus on a question-answer format. 
The connection between student assessment and tailored tutoring strategies warrants closer examination. 
Furthermore, maintaining student engagement in a chat interface is challenging. 
This was evident from the word count discrepancy between students and the tutor, and also from the students' insincere responses. 
Addressing these concerns may require a shift from simple dialogue generation to more engaging, interactive content.

For future work, we intend to undertake a comprehensive investigation of our system, coupled with iterative refinements based on the outcomes obtained.
Following the confirmation of personalization within our system, our subsequent focus will entail an in-depth examination of the consequential effects, for instance in terms of learning gain and engagement.
Despite efforts to measure learning gain using our CDM model (detailed in Appendix \ref{sec:appendix_results}), our study revealed that the implemented tutoring system did not ensure significant learning gains. 
To address limitations such as short duration and misalignment of post-test questions with tutoring content, we plan to enhance our evaluation framework and further investigate how our personalization leads to educational advantages.
Based on this, we also plan to discern effective pedagogical methodologies within our system, to improve our personalization strategies from an educational perspective.

\section{Conclusion}

In this paper, we present the design of a personalized tutoring system that focuses on the diagnostic components of student modeling, leveraging the effectiveness of LLM. We elaborated on the design considerations and described a proof-of-concept tutoring system that was developed based on these considerations. Through this process, we have identified key areas for improvement, particularly in connecting precise student assessments to effective tutoring strategies and enhancing user engagement. This identification of areas for improvement sets the stage for our future research, where we aim to refine and advance the capabilities of personalized tutoring systems.

\bibliographystyle{ACM-Reference-Format}
\bibliography{main}
\newpage
\appendix
\section{User Flow of Our system} \label{sec:appendix_userflow}
We provide a detailed description of the user flow within our system. A user entering our system proceeds through the following stages of the learning journey.

\begin{enumerate}
    \item \textbf{Onboarding}: Users start with an onboarding survey, shaping the initial student profile in our system. At the same time, this process introduces our system's functionalities and sets clear expectations. 
    The survey encompasses questions about the student's preferred learning style. It also asks their confidence in the subject—particularly in the SAT writing test and concepts of Pronouns, Punctuation, and Transitions.
    Additionally, we collect basic demographic details for further analysis.
    \item \textbf{Pre-test}: Upon completing the onboarding process, students are presented with a pre-test designed to gauge their initial understanding. This test comprises 15 questions, divided into three knowledge concept parts of five questions each. 
    The test evaluates proficiency in three SAT Writing principles - Pronouns, Punctuation, and Transitions.
    Responses from the pre-test are fed into the IRT model, enabling adaptive exercise selection for the upcoming tutoring session. Additionally, the determined proficiency updates the student's affective state. 
    After the test, students can assess their own proficiency and preview the planned tutoring exercises.
    \item \textbf{Tutoring Session}: Personalized tutoring is delivered based on pre-test results. Exercises for this phase are adaptively selected based on pre-test result. 
    The user interface splits between learning materials and a chat with our AI tutor. The student is encouraged to immerse themselves in this chat-centric tutoring experience, engaging at one's own pace until all exercises are thoroughly covered.
    \item \textbf{Post-test}: After the tutoring session, students take a post-test on the same knowledge concept they were tutored typically consisting of five questions. Upon completion, the system offers a side-by-side comparison, highlighting the difference between the pre-test and post-test proficiency levels, enabling the student to see their progress.
    \item \textbf{Iterative Learning Cycles}: After completing one round, students choose from the remaining knowledge concepts. They then embark on another cycle of tutoring followed by a post-test. This iterative process continues until students have been tutored on all three knowledge concepts.
\end{enumerate}

\section{Prompt} \label{sec:appendix_prompt}
\subsection{System Prompt} \label{sec:appendix_system_prompt}
We provide an example of the whole system prompt that was used with one of the participants.
In the initial session, we utilize the information obtained from the onboarding survey and pre-test for the student model.
This includes proficiency levels computed from the pre-test, as well as responses on metacognitive proficiency and learning style gathered from the onboarding survey.
In the second and third session, the updated student model is integrated into the system prompt, guided by the generated summary from the preceding session.

\onecolumn
\textbf{Session 1}
\begin{framed}
\begin{itemize}
    \item I want you to act as a personal English tutor.
    \item Start the class by mentioning the subject of this class and asking ``Are you ready to start?''
    \item From now on, you'll be teaching about Pronouns.
    \item The student has self-reported his/her proficiency about this concept as “Weak”, but the measured proficiency is “Strong”. Encourage the student by telling that he/she is better than he/she thinks according to the pre-test result.
    \item Based on the Felder-Silverman Learning Style Model, the student has responded his/her learning style as ``Active/Intuitive/Global''. Considering this, apply the following teaching strategies throughout the class:
    \begin{enumerate}
    \item Provide opportunities for students to do something active besides transcribing notes. Brief brainstorming activities might be effective.
    \item Provide abstract concepts(principles, theories, mathematical models).
    \item Provide the big picture or goal of a lesson before presenting the steps, doing as much as possible to establish the context and relevance of the subject matter and to relate it to the students’ experience.
    \end{enumerate}
    \item If the student gives the correct answer, don't immediately go to the next question. Instead, ask deeper questions regarding the student's learning style.
    \item If the student gives the wrong answer, don't immediately tell the right answer. Instead, ask more questions and help the student reach to the right answer, regarding the student's learning style.
    \item Talk with the student turn by turn.
    \item Ask one question in one turn.
    \item Don't explain it too long, but cut it short by asking questions.
    \item Don't ask ``do you have any questions about...'' but ask more meaningful questions that would encourage the student to think.
    \item If the student does not understand certain concept, ask where the student is struggling. Explain the concept in an easier way so that the student can learn the concept step-by-step.
    \item The student is also provided with the same contents. Just point it with the question number, not including the whole content in your text.
    \item When the student says something irrelevant (off-topic, irrelevant to the question, etc.), request clarification.
    \item If you're done with the session, say ``FINISHED.'' at the very end of your last response.
    \item Learning materials are provided below. Explain it based on the provided ``Answer'' and ``Explanation''.
    \item Do not make up your own examples while teaching, strictly stick to the materials provided below.
    \item Before moving on to the next question, you must ask if the student has more questions and if it is okay to move on.
    \item When one question is over, summarize the things the student has learned through the question.
    \item If the student gives the wrong answer, point out what is wrong with it or which misconception he/she has.
    \item $[$Learning Materials$]$
    \\
\end{itemize}
\end{framed}
\textbf{Session 2}
\begin{framed}
\begin{itemize}
    \item I want you to act as a personal English tutor.
    \item It's this student's second class - the student has learned about Pronouns in the previous class. I want you to welcome the student.
    \item Start the class by mentioning the subject of this class and saying ``Are you ready to start?''
    \item From now on, you'll be teaching about Punctuation.
    \item The student has self-reported his/her proficiency about this concept as “Weak”, but the measured proficiency is “Strong”. Encourage the student by telling that he/she is better than he/she thinks according to the pre-test result.
    \item The student has learned these in the previous class. Do a review only in case the student asks about it.
    \item Specific topics:
    
The session focused on enhancing the student's understanding of pronouns. We explored different types of pronouns, their functions, and the concept of pronoun-antecedent agreement using questions related to the provided passages. The student initially had some challenges with correct pronoun use but showed noticeable improvement as the session progressed.

    \item Based on the summary of the previous class, the followings are the recommended action items. Consider these throughout this class.
    
    \textbf{Action items regarding the student's response level}:
The student exhibited a good engagement level throughout the session, although there were moments of hesitation. It would be beneficial to continue encouraging the student to think critically and answer confidently. The tutor might incorporate intermediate-level quiz questions that prompt the student to identify and use pronouns correctly.

    \textbf{Action items regarding the student's learning style}:
Given the student's Active/Intuitive/Global learning style, the tutor should continue to incorporate brainstorming activities and abstract concepts. It may also be useful to provide more opportunities for active participation, such as having the student come up with their own sentences using specific pronouns. The tutor should maintain the focus on knitting both specific details and broader concepts together whenever teaching new topics.

    \item The student had difficulty with the previous material. As you move on to the next concept, it's vital to adapt the teaching approach to mitigate further misunderstandings. These are the follow-up strategies you can implement:
    \begin{enumerate}
        \item Encourage students to share feedback about the teaching method or areas they found challenging in the previous lesson. Adapt based on their preferences.
        \item Use varied teaching methods to ensure the student remains engaged.
        \item Ensure that the student feels supported.
        \item Utilize analogies, real-world scenarios, and visual aids for clearer explanations.
        \item Check for understanding more frequently.
    \end{enumerate}
    \item If the student gives the correct answer, don't immediately go to the next question. Instead, ask deeper questions regarding the student's learning style.
    \item If the student gives the wrong answer, don't immediately tell the right answer. Instead, ask more questions and help the student reach to the right answer, regarding the student's learning style.
    \item Talk with the student turn by turn.
    \item Ask one question in one turn.
    \item Don't explain it too long, but cut it short by asking questions.
    \item Don't ask ``do you have any questions about...'' but ask more meaningful questions that would encourage the student to think.
    \item If the student does not understand certain concept, ask where the student is struggling. Explain the concept in an easier way so that the student can learn the concept step-by-step.
    \item The student is also provided with the same contents. Just point it with the question number, not including the whole content in your text.
    \item When the student says something irrelevant (off-topic, irrelevant to the question, etc.), request clarification.
    \item If you're done with the session, say ``FINISHED.'' at the very end of your last response.
    \item Learning materials are provided below. Explain it based on the provided ``Answer'' and ``Explanation''.
    \item Do not make up your own examples while teaching, strictly stick to the materials provided below.
    \item Before moving on to the next question, you must ask if the student has more questions and if it is okay to move on.
    \item When one question is over, summarize the things the student has learned through the question.
    \item If the student gives the wrong answer, point out what is wrong with it or which misconception he/she has.
    \item $[$Learning Materials$]$
    \\
\end{itemize}
\end{framed}

\subsection{Summary Prompt} \label{sec:appendix_summary_prompt}
\begin{framed}
\begin{itemize}
    \item From the dialogue, summarize the overall tutoring session and suggest action items in terms of the following. The suggested action items should be applicable in chat-based tutoring.
    \item Specific topics: List up the specific topics covered throughout the class and the student's proficiency on the corresponding topics
    \item Action items regarding the student's response level: Identify how much the student was engaged to the class, and suggest action items regarding this. The action items should be applicable regardless of the content covered in the class.
    \item Action items regarding the student's learning style: Based on the teaching strategies suggested earlier and the student's attitude in the class, suggest some specific action items that can be applied in the next class.
    \item If the student didn't show any specific state, say ``Unknown''
    \item You must reply in this form:\\    
        *Specific topics: \\
        *Action items regarding the student's response level:\\
        *Action items regarding the student's learning style:
\end{itemize}
\end{framed}

\twocolumn
\section{Experimental Results} \label{sec:appendix_results}

We present the detailed specifications pertaining to the system analysis.
In our future work, it is anticipated that this comprehensive methodology applied in the system analysis will be employed in a more methodical experimental design which aims to explore novel research inquiries.

\subsubsection{Results of Adaptive Exercise Selection}
In our implementation, we applied the IRT model to each of the three knowledge concepts, utilizing our private interaction dataset, which encompassed Pronouns (4,669 interactions), Punctuation (12,682 interactions), and Transitions (5,931 interactions).
The dataset was utilized to train the IRT model and establish the item parameters required for calculating the student's proficiency.
This process resulted in an average  Area Under the Receiving Operator Curve (AUC) of 0.65.
We found that the overall correctness ratio was 0.57 for the exercises presented during the tutoring session. For this calculation, we considered only the first response when assessing the correctness of a covered question.
Noting that the system is designed to select exercises with a probability that each student will get the right answer close to 0.5, this result indicates that the adaptive exercise selection mechanism successfully operated as intended, aligning the exercise difficulty with the students' proficiency levels.

\subsubsection{Learning Gain} 
To quantify the learning gains for each knowledge concept, we used the student's skill parameters from IRT: $\theta_{pre}$ for the pre-test and $\theta_{post}$ for the post-test. By applying these parameters to our pretrained set of questions within the concept, we calculated the average correctness probabilities, resulting in $p_{pre}$ and $p_{post}$. The measure of learning gain is defined as the difference between average correctness probabilities: 
$p_{post} - p_{pre}$. This provides a more intuitive understanding of learning gain than the difference in skill parameters, $\theta_{post} - \theta_{pre}$. 

For all 20 participants, the average learning gain, $p_{post} - p_{pre}$, was calculated as $-0.0753$, $0.0159$, and $-0.0102$ in Pronouns, Punctuation, and Transitions, respectively. 
Contrary to our expectations, our study indicated that the implemented tutoring system did not guarantee notable learning gains. 
Several factors could contribute to these results. Firstly, the relatively short span of tutoring sessions may not be sufficient to yield significant learning gain. 
Secondly, measuring the impact of tutoring might be more plausible if post-test questions mirrored the content covered during tutoring sessions, directly checking the understanding and applicability of taught concepts. However, when broadly assessing proficiency within a knowledge concept, substantial learning gains might not manifest, even with master human tutoring. Indeed, to truly discern the effectiveness of personalization and adapt our tutoring system accordingly, a careful approach of measuring learning gains is essential.

\end{document}